\documentclass[prl,twocolumn]{revtex4}
\usepackage{graphicx}
\usepackage{amsmath}
\usepackage{epsfig}

\begin{document}

\title{All-optical delay of images using slow light}

\author{Ryan M. Camacho, Curtis J. Broadbent, Irfan Ali-Khan, John C. Howell}
\affiliation{Department of Physics and Astronomy, University of
Rochester, Rochester, NY 14627, USA }

\begin{abstract}
Two-dimensional images carried by optical pulses (2ns) are delayed
by up to 10 ns in a 10 cm cesium vapor cell. By interfering the
delayed images with a local oscillator, the transverse phase and
amplitude profiles of the images are shown to be preserved. It is
further shown that delayed images can be well-preserved even at very
low light levels, where each pulse contains on average less than one
photon.
\end{abstract}

\maketitle \

All-optical methods for delaying images may have great potential in
image processing, holography, optical pattern correlation, remote
sensing, and quantum information.  For example, in many digital
image processing applications, the amplitude and phase information
of images must be preserved. Electronic conversion of optical images
requires relatively intense optical fields and information is lost
in analog to digital conversions. Alternatively, one could use a
long free-space delay line; however, diffraction and physical space
limitations impose serious restrictions on such a system.  A small
all-optical buffer in which the phase and amplitudes of the image
are preserved would solve these problems.

In this work, we report on a series of several ``slow light"
experiments showing that two-dimensional images can be delayed while
preserving the amplitude and phase information of the images.  The
system we use has several noteworthy characteristics. It requires no
additional laser beams to prepare the slow light medium. This
results in low background noise and a high signal-to-noise ratio in
the delayed image, even at very low light levels.   The transverse
images can be delayed by many times the pulse length without
affecting the phase stability of the image. This is demonstrated by
interfering the images with a pulsed local oscillator and monitoring
the interference pattern.  The interference stability has almost no
dependence on fluctuations in the group velocity in the slow light
system, but only on the phase velocity, which is unaffected by a
slow light medium. This property leads to stable and high fringe
visibility when the delayed image interferes with a local oscillator
even if the slow light medium has moderate thermal instabilities.

Slow light is the name given to the subfield of optics that deals
with the slow group velocity (the velocity at which the energy
travels) of a light pulse in a highly dispersive
medium\cite{boyd02,chiao02}. A dispersive medium has  a
frequency-dependent index of refraction, which occurs in any system
having frequency-dependent absorption. The slow light medium used in
the present experiments is a hot Cesium (Cs) vapor. Delay bandwidth
products (the delay of the light signal in the medium multiplied by
the bandwidth of the signal) in excess of 50 \cite{camacho06} can be
achieved by spectrally tuning the signal between the two D$_2$
hyperfine ground states.

While several methods for achieving slow light have been explored,
the system discussed here is particularly attractive. For example,
most slow light systems, such as electromagnetically induced
transparency
\cite{kasapi95,jain95,kash99,Budker99,hau99,liu01,turukhin02},
coherent population oscillations
\cite{bigelow03,Zhao05,palinginis05} or spectral
hole-burning\cite{camacho06_2}, require additional light fields to
prepare the medium.  The use of additional light fields results in
transverse spatial inhomogeneities in the group velocity of the
medium.  To the authors' knowledge, the only previous studies of
transverse images in a slow light medium were performed by Harris'
group \cite{kasapi95, jain95}. In the Cs system used here, the group
velocity is the same in all directions. Also, this system has
relatively low loss and minimal broadening of the pulse.

A brief overview of some of the most important theoretical elements
will be given.  A more detailed treatment of the propagation of a
Gaussian light pulse through a medium with two widely spaced
absorbing Lorentzian optical resonances may be found in previous
work \cite{tanaka03, macke06,camacho06,zhu06}. Assuming two
Lorentzian absorption resonances of equal strength, the complex
index of refraction may be written as the sum of a free-space term
(unity) and two distinct Lorentzian terms:
\begin{equation}
 n(\delta) = 1+\frac{\beta}{2}\left( \frac{1}{-(\delta+\omega_0)-i\gamma} +
 \frac{1}{-(\delta-\omega_0)-i\gamma}\right ),
\end{equation}
where each resonance has a strength  $\beta$, a spectral half-width
$\gamma$, $\delta$ is the detuning from the midpoint between the two
resonances, and 2$\omega_0$ is the separation between the
resonances. If the bandwidth of the input pulse is much less than
the resonance spacing 2$\omega_0$, we may expand the refractive
index in a power series and keep only the first few terms.  The real
and imaginary parts of the index of refraction, responsible for
pulse delay and absorption respectively, may then be written as

\begin{subequations}
\begin{equation}
n' \approx 1 + \frac{\beta}{\omega_0^2} \delta +
\frac{\beta}{\omega_0^4} \delta^3
\end{equation}
\begin{equation}
n'' \approx \frac{\beta \gamma}{\omega_0^2}+ 3 \frac{\beta
\gamma}{\omega_0^4} \delta^2.
\end{equation}
\end{subequations}

It can be seen that $dn'/d\delta = n''/\gamma$, which can be used to
obtain a simple form for the group velocity.   Combining this result
with $\alpha=2\omega n''/c$, where $\alpha$  is the optical
intensity absorption coefficient of the medium at the pulse carrier
frequency, one obtains an approximate group velocity
\begin{equation}
v_g \approx \frac{c}{\omega \frac{dn'}{d \delta}} = \frac{c
\gamma}{\omega n''} = \frac{2 \gamma}{\alpha}.
\end{equation}

If the medium is of length \emph{L}, the resulting pulse delay in
the medium is approximately
\begin{equation}
\tau_g = \frac{L}{v_g} \approx \frac{\alpha L}{2 \gamma}.
\end{equation}

Several points are worth noting.  First, the pulse delay is given
entirely by the width of the optical resonances and the optical
depth of the medium at the pulse center frequency, and is
independent of the separation of the optical resonances.    Second,
the signal bandwidth is only limited by the frequency separation of
the two optical resonances (e.g., 9.2 GHz in Cs), which can be very
large. Third, the group velocity is isotropic in the sample. Fourth,
while not treated here, the temporal broadening of the pulse in the
delay medium is relatively small \cite{camacho06}, owing to the fact
that the dispersive broadening dominates the absorptive broadening.

\begin{figure}
\centerline{\includegraphics[scale=.4]{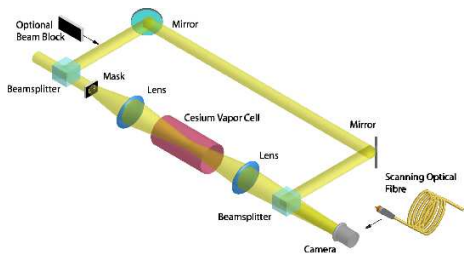}}
\caption{Experimental setup for the delay of transverse images.
Light pulses of 2 ns duration are incident on a 50:50 beamsplitter.
The transmitted pulses then pass through an amplitude mask and a 4f
imaging system. The transmitted and reflected pulses are recombined
at another 50:50 beamsplitter. The transmitted part traverses a path
approximately 5 feet shorter than the reflected path, and arrives at
the second beamsplitter about 5 ns sooner than the reflected pulse,
preventing interference between the two pulses. The temperature of
cesium vapor can then be adjusted to give 5 ns of delay, resulting
in interference.  In the low-light-level experiment, the pulses are
attenuated such that each pulse contains on average less than one
photon and the reflected path is blocked. A scanning optical fiber
is used to collect the photons in the image plane and the photon
arrival times recorded using a photon counter with time-to-digital
converter.} \label{experiment}
\end{figure}

Consider the experimental setup represented in Fig. 1.   Light
pulses with a duration of 2~ns full width half maximum (FWHM),
repeating every 7~ns, are used in the experiment.  The pulses are
generated by passing a CW laser beam through a fiber-coupled,
high-speed electro-optic modulator (DC to 16 Gb/s). The laser
frequency is set halfway between the two hyperfine ground state
resonances of the D$_{2}$ lines (852 nm wavelength) in Cesium.  The
pulses enter an unbalanced Mach-Zehnder interferometer with a
free-space path mismatch of 5~ns. The pulses propagating in the long
path are reference pulses (local oscillator) that are made to
interfere with the pulses exiting from the slow light medium in the
short path.  In the short path, the pulses impinge on an amplitude
mask, and are called image pulses. A 4.5-lines-per-millimeter test
pattern was used as the mask. The hot Cs vapor is in the middle of a
4f imaging system, which consists of two identical 150 mm lenses. In
a 4f imaging system the object is placed a focal length (150 mm) in
front of the first lens, the distance between the two lenses is two
focal lengths (300 mm), and the image is produced in the back focal
plane of the second lens.  The 4f system was used to eliminate the
quadratic phase in the image plane. The group delays in the cell are
varied by changing the vapor pressure through temperature control of
the cell.

The image and reference pulses interfere via the second 50/50 beam
splitter.  One of the mirrors in the long path has a piezo-actuated
mount allowing for precision translations of the mirror.  Movements
on the order of a few nm are possible allowing for control of the
relative phase of the reference and image pulses at the
beamsplitter. By translating the mirror through a phase shift of
$\pi$ radians, it is possible to measure the fringe visibility. The
interference images were measured on a CCD camera run in continuous
mode (a CCD camera capable of gating pulses in a 2 ns pulse window
was unavailable for the present study).

The experimental apparatus for the weak light fields experiment is
considerably different from the macroscopic light fields experiment
discussed above.  Pulses of light of 4 ns FWHM duration, repeating
every 330 ns, are created in the same way as the macroscopic light
fields. However, the pulses are attenuated so that, on average,
there is less than 1 photon per pulse impinging on the amplitude
mask. The long arm of the Mach-Zehnder interferometer is blocked,
leaving only a straightforward 4f imaging system and the slow light
medium. To recreate the image, a scanning multimode fiber with a 62
$\mu$m diameter core is used to collect the photons in the image
plane. The multimode fiber is coupled to a single-photon counting
module with 300~ps detector jitter (Perkin Elmer SPCM). The
electronic signal from the detector is sent to a 16 ps resolution
time-to-digital converter and is time-stamped.  The multimode fiber
is continuously scanned using computer-controlled translation stages
with 20~nm resolution. The position of the the translation stages is
recorded as a function of time.  The clock of the
computer-controlled translation stages is synchronized with that of
the time-to-digital converter using the electronic pulse that is
driving the electro-optic modulator.  Thus, the 2-dimensional image
is reproduced by binning the photon detection events into the
2-dimensional positions at which they were detected. Background
counts (e.g., light from the room or detector dark counts) in the
image are significantly reduced by only accepting time-binned data
centered around the relative delay of the image, within a time
window that is determined by the parameters of the pulse. This can
be done in postprocessing of the image by looking at the temporal
histogram of arrival times.  The interferometer was not used in the
weak fields experiment for the practical reason that the relatively
large interferometer was not phase stable for the entire scan
duration.

\begin{figure}
\centerline{\includegraphics[scale=.28]{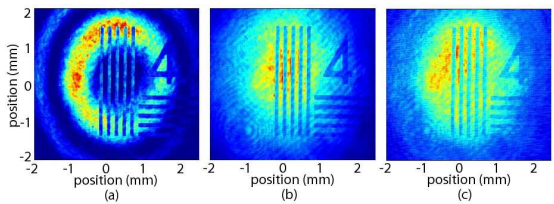}}
\caption{Interference of a delayed image with a slightly diverging
local oscillator. (a) An image (a black pattern of bars and a
numeral) delayed by 5 ns interferes with a reference beam and
produces a ring pattern superimposed with the image.  In the central
dark spot, the two beams destructively interfere and cancel one
another except in the image, which remains relatively bright. In the
ring surrounding the central spot, the two beams constructively
interfere and add to create a bright ring except in the image, which
remains relatively dim.  The succeeding rings alternate between
constructive and destructive interference. (b) and (c) show the same
superposition of the two beams, but in the absence of slow light. In
(b), the wavelength of the laser is tuned outside of the dispersive
region and in (c) the cesium cell is removed.  In both cases, no
interference between the beams can be seen. } \label{interference}
\end{figure}

Consider the results for the macroscopic image interference shown in
Fig. \ref{interference}. In Fig. \ref{interference}(a), the Cs cell
temperature is set to give 5 ns of delay, which matches the arrival
time of the image pulses at the second beamsplitter to that of the
reference pulses. The situation in which both pulses arrive at the
beamsplitter simultaneously will be referred to as``temporally
matched".  The intensity along the two paths is balanced for maximum
interference. The phase of the local oscillator is set to give a
dark fringe in the center of the image. Several $\pi$ radians of
phase shift across the image can be observed. The only regions that
do not experience interference are the image points of the dark
patterns of the amplitude mask. Since there is no light in the
delayed image at those points, the local oscillator creates a
constant background where the dark regions of the amplitude mask are
imaged.  Hence, at the center of the dark fringe the inverse image
is created. An interference visibility of 90 \% $\pm$ 1 \% was
observed for the temporally matched pulse regime.  The pulses from
the two arms of the Mach-Zehnder interferometer are then misaligned
in time, so as to arrive at the beamsplitter at different times.
This is accomplished by either removing the cell or by tuning the
delay of the pulse. In both cases the observed visibility dropped,
as seen in Figs. 2 b) and c), respectively. The images show the same
number of phase shifts as the temporally matched pulses but the
interference visibility (after balancing the intensity in each arm)
is 6\% for cell removal and 15\% for delay tuning, far lower than
the 90\% visibility for the temporally matched case. As a note,
there is always a small amount of CW light leaking through the
electro-optic modulator, which has a 100:1 extinction ratio. The CW
light is the primary culprit in giving the nonvanishing interference
visibility when the pulses are temporally mismatched. The amount of
CW light can be much greater than 1\% of the total light since it is
constantly ``on", which can lead to a much larger integrated CW
signal.  In the interference experiment, the CW background is about
5\%. The CW background light can be removed by using a camera that
is able to gate around a 2 ns window in a fashion similar to that of
the low light-level experiment.

\begin{figure}[b]
\centerline{\includegraphics[scale=.4]{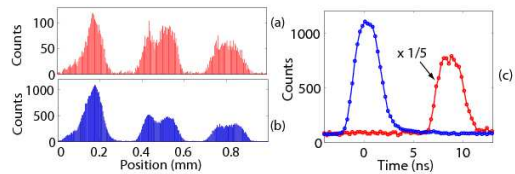}} \caption{(a)
Delayed and (b) non-delayed one-dimensional low-light-level image
with (c) accompanying histograms of photon arrival times. Each pulse
contains, on average, 0.5 photons before striking the image mask.}
\label{1dimage}
\end{figure}

\begin{figure}
\centerline{\includegraphics[scale=.5]{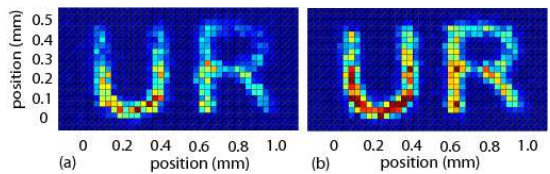}} \caption{False
color representation of a (a) Delayed and (b) non-delayed
two-dimensional low-light-level image.  An optical fiber was
raster-scanned across a two dimensional image consisting of the
letters ``UR". Though attenuated, the delayed imaged shows similar
image fidelity and resolution to the non-delayed image.  Each pulse
contains, on average, 0.8 photons before striking the image mask.}
\label{2dimage}
\end{figure}

\begin{figure}
\centerline{\includegraphics[scale=.30]{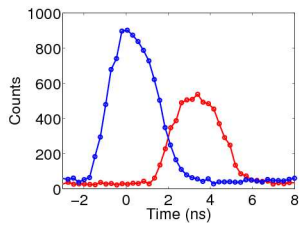}}
\caption{Histogram of photon arrival times showing the delayed (red)
and non-delayed (blue) two-dimensional image shown in Fig. 5}
\label{2ddelay}
\end{figure}

The experimental results for the weak field images are shown in
figures \ref{1dimage} through \ref{2ddelay}.  Figs. \ref{1dimage}(a)
and (b) show a delayed and non-delayed one-dimensional image (a bar
test pattern) where each pulse impinging on the image mask contains,
on average, 0.5 photons.  The images are measured by scanning an
optical fiber in a line across the image plane for a total duration
of 36 seconds. A histogram of the photon arrival times is made for
each incremented position of the fiber (an effective pixel) as it
scans across the image (shown in Fig. \ref{1dimage}(c)). The
measured image is the convolution of the image with the fiber core.
For these scans, the laser frequency is set halfway between the
optical resonances and the temperature of the cell is set to give 9
ns of delay (shown in red). The process is repeated but with the
laser frequency tuned far from either resonance, which gives almost
no delay (shown in blue). Approximately 99\% of extraneous counts
from background light and detector dark counts are removed by
constructing the images using only those photons which arrive in a 4
ns time window (out of the entire 330 ns window) centered on the
middle of the pulse arrival time distribution.  An analysis of the
undesirable counts led to an estimate of approximately 2 extraneous
counts per spatial bin shown in Fig. \ref{1dimage}, which is in good
agreement with the image noise.

Figure \ref{2dimage} shows the delay of a two-dimensional image
comprised of the letters ``UR"  representing the researchers'
institution.   In this part of the experiment, each pulse contains,
on average, 0.8 photons before arriving at the image mask.  The
image is constructed by raster scanning a fiber across the image
plane in a total time of approximately 48 seconds. The time-binned
filtering technique described above was also used to remove
background counts from the two-dimensional images.   A histogram of
the photon arrival times for the two dimensional images of Fig.
\ref{2dimage} is shown in Fig. \ref{2ddelay}., showing the delayed
image pulses. Even though every photon used to construct the image
is delayed by approximately 3 ns, the image is preserved with high
fidelity.

A few comments about the results are in order.  First, the
propagation through the medium is a classical effect, meaning that
its behavior does not change in going from classical fields to
quantum fields. A formal demonstration of the preservation of
quantum fields was not undertaken in the present study.  However,
the preservation of amplitude and phase as well as the low noise
characteristics imply that this system can be an integral part of
quantum image buffering.  The development of a highly multimode
quantum image buffer is a much different goal than that of the
preservation of two state systems that have been recently studied
(qubits)\cite{liu01,phillips01,Kocharovskaya01,turukhin02,bajcsy03,Yanikdec04}.
Second, the homogeneous linewidth of the cesium atoms ultimately
determines the upper limit in the absolute delay of our slow light
system. However, much narrower resonances could achieve a much
larger upper limit of the delay at the expense of the usable signal
bandwidth.

In conclusion, it is shown that a transverse image can be delayed in
a slow light buffer.  The buffer is shown to be able to delay the
image by many pulse widths, while also preserving its amplitude and
phase characteristics. The image is interfered with a pulsed local
oscillator. When the local oscillator and image pulse are temporally
overlapped, high visibility fringes with 90\% visibility are
observed, demonstrating the preservation of phase information after
a 5 ns pulse delay. When the local oscillator and image pulse are
temporally misaligned, low visibility fringes are observed,
demonstrating the pulsed nature of the imaging system. The slow
light system is then used to delay an image using weak-light pulses.
Pulses with less than one photon on average are used to image an
amplitude mask. The image of the mask is reproduced with high
fidelity and low noise, demonstrating 9 ns pulse delays of images at
very weak light levels.

This research was supported by DOD PECASE, DARPA Slow Light, NSF,
and the Quantum Imaging MURI.

\bibliographystyle{slowbst}
\bibliography{slowbib}

\end{document}